\documentclass[twocolumn,aps,prc,showpacs,preprintnumbers,amsmath,amssymb,10pt]{revtex4-1}

\usepackage{graphicx}
\usepackage{dcolumn}
\usepackage{setspace}
\usepackage{bm}
\usepackage{multirow}
\usepackage[dvipsnames]{xcolor}

\begin{document}
\title{Influence of additional neutrons on the fusion cross-section beyond the N=8 shell}

\author{S. Hudan}
\author{H. Desilets}
\author{Rohit Kumar}
\author{R.~T. deSouza}
\email{desouza@indiana.edu}
\affiliation{%
Department of Chemistry and Center for Exploration of Energy and Matter, Indiana University\\
2401 Milo B. Sampson Lane, Bloomington, Indiana 47408, USA}%

\author{C. Ciampi}
\author{A. Chbihi}
\affiliation{GANIL, CEA/DRF-CNRS/IN2P3, \\
Blvd. Henri Becquerel, F-14076, Caen, France} %

\author{K.~W. Brown}
\affiliation{Facility for Rare Isotope Beams and Department of Chemistry, Michigan State University, East Lansing, MI 48823, USA} %

\date{\today}
\begin{abstract}
Fusion enhancement for neutron-rich isotopes of oxygen on carbon nuclei was probed. To measure the fusion cross-section
a $^{20}$O beam accelerated to E$_{lab}$/A=2.7 MeV bombarded the active-target detector MuSIC@Indiana with a fill gas of CH$_4$.  Examination of the average fusion cross-section over the interval 12 MeV $\leq$E$_{c.m.}$$\leq$ 17 MeV for $^{16-20}$O + $^{12}$C reveals that while even isotopes of oxygen exhibit essentially the same cross-section, the cross-section for odd isotopes can be either enhanced or suppressed relative to the even A members of the isotopic chain. Theoretical models fail to explain the observed experimental results.
\end{abstract}


\maketitle

The nature of neutron-rich nucleonic matter, both its structure as well as the reactions it undergoes, is of fundamental interest to the fields of both nuclear physics and nuclear astrophysics. Recent discovery of neutron star mergers as a nucleosynthetic site for heavy elements \cite{Nicholl17} underscores the importance of characterizing the behavior of extremely asymmetric nuclear matter. For reactions involving neutron-rich nuclei the central question is: “Do neutron-rich nuclei exhibit different behavior from more stable nuclei?” For increasingly neutron-rich nuclei approaching the neutron drip-line, one can expect a change based upon the increased importance of coupling to continuum states \cite{Casal20, Michel08}, emergence of new collective modes \cite{Nakatsuka17}, and changes to pairing/pairing dynamics \cite{Magierski17, Scamps19}.  These changes can result in an enhancement of the fusion cross-section \cite{Hagino21} as well as breakup and transfer \cite{Cortes20}. Systematic experimental investigation of these topics for neutron-rich nuclei can provide insight into how nuclear reaction properties evolve with isospin, a central topic of next generation radioactive beam facilities \cite{FRIB, RIKEN, GANIL}.

Reactions with light nuclei present a unique opportunity to examine the character of extremely neutron-rich matter close to the drip-line and explore the impact on fusion, breakup, and transfer. The systematic study of heavy-ion fusion at near-barrier energies for an isotopic chain is a powerful tool to investigate neutron-rich matter at low-density probing the neutron and proton density distributions and how they evolve as the two nuclei approach and overlap \cite{Gasques07, Steinbach14a, Singh17, Vadas18, Hudan20}.  At high energies, measurement of the interaction cross-section revealed that density distributions were spatially extended resulting in the discovery of halo nuclei \cite{Tanihata85a,Tanihata85b}. Fusion of neutron-rich nuclei at low energies probes both the increase in the spatial extent of the ground-state density distribution, as well as changes in dynamics as the system fuses. The density distribution probed by the fusion cross-section is not just the one-body density distribution but includes the quantal structure effects for the fusing system. The same interplay of attractive and repulsive forces in the two-component nucleonic matter that governs the composition and spatial extent of the low-density tails also impacts the collision dynamics.

Comparison of the average fusion cross-section measured in $^{12-15}$C + $^{12}$C \cite{Carnelli15, Calderon15} with neutron-excess clearly indicates an increased cross-section for $^{15}$C \cite{deSouza21}. Neither static nor dynamical models are able to describe this unexpected increase in the average fusion cross-section for $^{15}$C, motivating investigation of isotopic chains for other light nuclei.
Fusion of neutron-rich oxygen nuclei provides a good opportunity to investigate this topic further and specifically explore the impact of additional neutrons beyond the N=8 shell. Using radioactive beam facilities 
one can now explore fusion of neutron-rich oxygen beams to A=22 -- nearly to the last neutron-bound nucleus $^{24}$O \cite{GANIL, FRIB}. 
Closure of the 1p$_{1/2}$ shell with N=8 means that all isotopes with A$>$16 have their valence neutrons in the {\em{sd}} shell in their ground state.
Prior measurements revealed that for $^{19}$O + $^{12}$C a significant increase in the above-barrier fusion cross-section was observed as compared to $^{18}$O \cite{Singh17}. 
Well above the systematic increase expected, this increase was interpreted as an extended tail of the neutron density distribution as the system fuses \cite{Hudan20}. 
Whether more neutron-rich isotopes of oxygen also manifest an increased cross-section depends on the binding of the valence neutrons and the polarizability of the density distribution. 
On general grounds, pairing of the valence neutrons at the saddle point can impact this result.  
The present experiment provides the first measurement of the total fusion excitation function for $^{20}$O + $^{12}$C and compares the average above-barrier cross-section with that of less neutron-rich isotopes.

\begin{figure}[h]
\begin{center}
\includegraphics[scale=0.350]{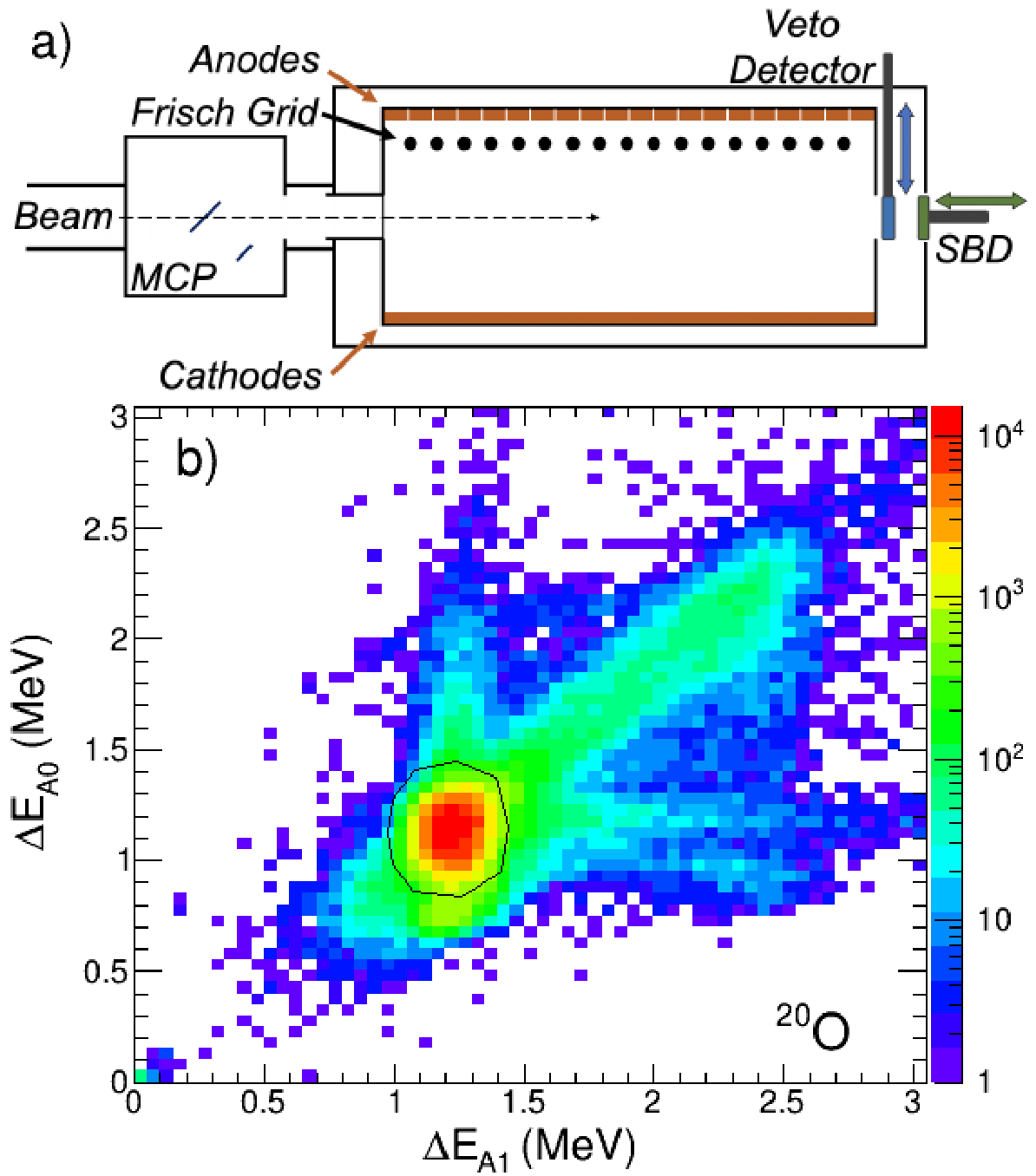}
\caption{
Panel a: Schematic of the experimental setup used to measure fusion of $^{20}$O + $^{12}$C. The tilted foil MCP detector, MuSIC@Indiana, the veto detector, and the downstream surface barrier silicon detector (SBD) are indicated. Panel b: Identification of the incident beam utilizing the energy deposit in the first two anodes of MuSIC@Indiana.
}
\label{fig:E831_setup}
\end{center}
\end{figure}

The experiment was conducted using the SPIRAL1 facility at the GANIL accelerator complex in Caen, France. 
A primary beam of $^{22}$Ne at E/A = 80.3 MeV bombarded a graphite target to produce a beam of $^{20}$O. After acceleration to an energy of E/A = 2.7 MeV by the CIME cyclotron, the $^{20}$O beam was selected in B$\rho$ by the ALPHA spectrometer and transported to the experimental setup.
A schematic of the experimental setup is shown in Fig.~\ref{fig:E831_setup}a.  
The beam passed first through a thin $\approx$35 $\mu$g/cm$^2$ aluminized mylar foil oriented at an angle of 45 degrees to the beam direction. 
Electrons ejected from the foil by the beam were accelerated and transported by an electromagnetic field \cite{Shapira00} onto the surface of a 2D position-sensitive microchannel plate (MCP) detector. 
The MCP detector provided a continuous measure of the beam position, size, and intensity during the experiment. Following the tilted-foil MCP the beam impinged on the active target detector MuSIC@Indiana at an intensity up to $\approx$3$\times$10$^4$ ions/s. 

The MuSIC approach, in which the detector gas in a transverse-field, Frisch-gridded ionization chamber  serves as both target and detection medium, has a couple of intrinsic advantages over thin-target measurements. Use of a MuSIC detector provides a direct energy and angle-integrated measure of the fusion products and allows simultaneous measurement of multiple points on the excitation function \cite{Carnelli15}. In addition, MuSIC detectors are self-normalizing since the incident beam is detected by the same detector as the reaction products. These advantages make MuSIC detectors a particularly effective means for measuring fusion excitation functions when available beam intensities are low \cite{Carnelli14, Johnstone21}. 

In this experiment, MuSIC@Indiana \cite{Johnstone21}, operated with a fill gas of CH$_4$ gas (99.99$\%$ purity), was utilized to measure the fusion cross-section. The purity of the gas was independently assessed {\em in situ} with a residual gas analyzer (MKS Microvision 2). During the experiment, in order to span the desired region of the excitation function, measurements were conducted at a sequence of pressures between 110 torr and 124 torr.  The gas volume is separated from the high vacuum  preceding it ($\approx$ 7$\times$10$^{-7}$ torr) by a 25 mm diameter, 2.6 $\mu$m thick mylar foil.
Passage of beam through the detector gas results in ionization characteristic of the incident ions as they traverse the detector. 
Segmentation of the anode into twenty strips along the beam direction, each 12.5 mm wide, yields a direct measure of this ionization trace. 
This segmentation of the anode plane means fusion events are associated with discrete locations (and therefore discrete energies) inside the detector.
If fusion occurs at a particular location a considerably larger ionization signal is observed due to the larger specific ionization of the fusion product (evaporation residue). To clearly identify that fusion has occurred the evaporation residue is required to stop in the active volume which requires up to six anodes.
Determination of the position at which the fusion occurs enables measurement of the fusion excitation function \cite{Carnelli14, Johnstone21}. 
To ensure a short collection time of the primary ionization produced by an incident ion, MuSIC@Indina was operated at
a reduced electric field of $\sim$0.7 kV/cm/atm in the active volume.
This field results in an electron drift velocity of $\sim$10 cm/$\mu$s for 
CH$_{4}$ \cite{Foreman81}, in principle allowing operation at beam intensities as high as 1$\times$10$^5$ ions/s.
Further details on the design, operation, and performance of MuSIC@Indiana have been previously  published \cite{Johnstone21, Johnstone22}.

For each ion, the first two anodes of MuSIC@Indiana were used to identify the ion using a $\Delta$E1-$\Delta$E2 measurement and ensure the absence of beam contaminants. 
Since upto six anodes were used to identify that fusion has occurred, for a fixed incident energy and gas pressure, thirteen points on the fusion excitation function were extracted. 
A key feature of MuSIC@Indiana is the ability to precisely insert a small silicon surface barrier detector (SBD) into the active volume from downstream, allowing accurate determination of the energy loss of different ions in the gas. The calibrated SBD was precisely inserted into the active gas volume under remote control with an accuracy of $<$0.1 mm and used to measure the beam energy at each anode. This measurement was performed for each gas pressure utilized.
Calibration with different beams of ions \cite{Johnstone21} eliminates the sensitivity to energy loss calculations which have uncertainties as large as 15$\%$. 

Readout of MuSIC@Indiana by the data acquisition system (DAQ) was triggered using a fast signal from the MuSIC@Indiana cathodes. The large amplitude of the signal associated with the deposit of several tens of MeV into the active volume made triggering the detector simple. 
For the majority of the $^{20}$O data, MuSIC@Indiana was operated in self-triggering mode. In this mode the data acquired is self-normalizing. For a small fraction of the data, in order to improve the data acquisition readout live-time (typically $\sim$70$\%$ for 1$\times$10$^4$ ions/s in self-triggering mode), a small veto  detector was inserted into the beam path downstream of the active volume of MuSIC@Indiana but within the same gas volume. 
This axial-field ionization chamber consisted of three $\sim$35 $\mu$g/cm$^2$ aluminized mylar foils spaced by 6 mm with a central anode. It provided a fast veto signal for the DAQ from un-reacted beam that exited the active volume of MuSIC@Indiana. 
Use of the veto allowed operation of the MuSIC@Indiana at a rate of 3$\times$10$^4$ ions/s, three times faster than without the veto detector, with a live time of $\approx$70$\%$. 
When the veto detector was employed, downscaled beam was acquired to provide normalization. The fusion cross-section was calculated using the gated $\Delta$E1-$\Delta$E2 spectrum and the downscale factor. Comparison of the fusion cross-section acquired with and without the veto detector confirmed that the measured cross-section was independent of its use.

Presented in Fig.~\ref{fig:E831_setup}b is the $\Delta$E$_{A0}$-$\Delta$E$_{A1}$ spectrum of MuSIC@Indiana used to identify an incoming beam particle.
Clearly evident is a single peak for $^{20}$O without discernible contaminant peaks. A locus of points with a slope of approximately unity corresponds to pileup events of two beam particles in MuSIC@Indiana. Near horizontal and vertical bands correspond to pileup events associated with scattering from the mylar window as the beam enters the active volume.
All analyzed events are tagged by requiring the identification condition indicated by the gate shown in Fig.~\ref{fig:E831_setup}b. This selection provides a count of the cleanly identified $^{20}$O ions incident on the detector.

\begin{figure}
\includegraphics[scale=0.42]{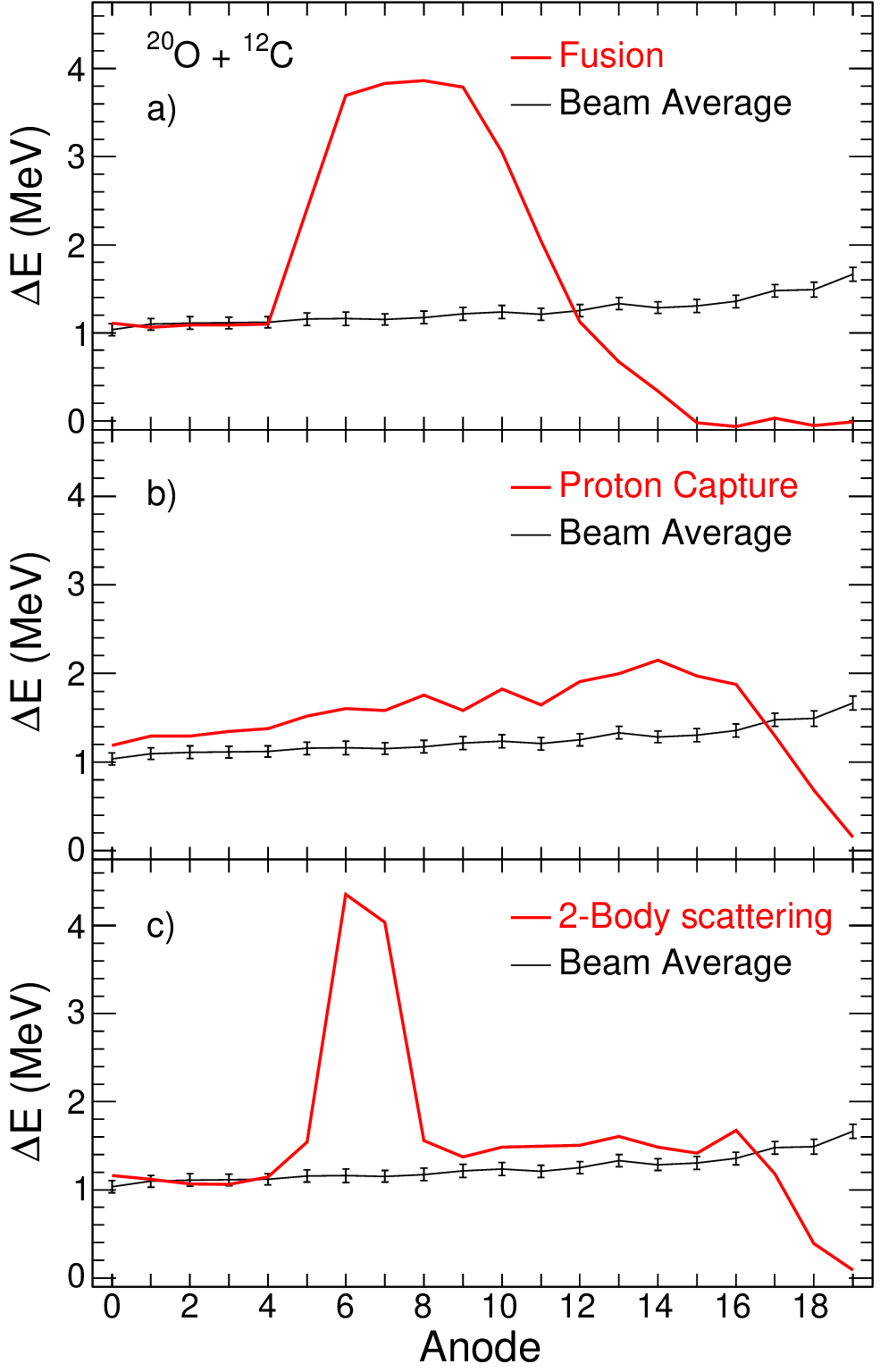}
\caption{ 
Representative traces from MuSIC@Indiana for different types of events. Indicated as the black line in all panels is the average trace for the $^{20}$O beam. Error bars indicate the standard deviation of the energy distribution for the beam on each anode. a) fusion in anode 4 b) Proton capture on $^1$H early in MuSIC@Indiana and c) a two-body event at anode 4.
}
\label{fig:Traces}
\end{figure}

Presented in Fig.~\ref{fig:Traces} are representative ionization traces measured in MuSIC@Indiana. 
The black line in all panels represents the average energy deposit of an $^{20}$O ion as it traverses the detector. Error bars indicate the standard deviation of the energy deposit in an anode. 
Indicated in panel a) is the ionization trace for a fusion event occurring in anode 5. Fusion of the $^{20}$O with $^{12}$C produces a $^{32}$Si with an excitation energy of $\approx $30-50 MeV. 
De-excitation of this compound nucleus via emission of neutrons, protons, and $\alpha$ particles results in an evaporation residue which, due principally to its larger atomic number as compared to the beam, manifests a larger ionization signal. 
Prior to fusion the incident $^{20}$O ion loses energy as it traverses the gas. Knowledge of the position at which the fusion occurs thus allows determination of the energy at which the incident ion fuses. Fusion events are assumed to occur in the middle of the anode in which they are detected. This simple assumption is reasonable as the beam does not stop in MuSIC@Indiana \cite{Johnstone22}.
Accurate determination of the fusion cross-section relies on the isolation of fusion events by their characteristic traces from competing processes. In addition to the veto detector unreacted beam is rejected by the presence of an appreciable signal (E$_{max}$$>$ 1 MeV) in the most downstream anode of MuSIC@Indiana. 
Following beam rejection, the principal contributions, aside from fusion, are from proton capture events or two-body (elastic) scattering. 
A representative trace for proton capture from the mylar window or early in the CH$_4$ gas is depicted in Fig.~\ref{fig:Traces}b. Clearly evident is the increased ionization due to the increased atomic number from $^{20}$O to $^{21}$F. Portrayed in Fig.~\ref{fig:Traces}c is a representative trace of a two-body event. 
In this event the incident ion is scattered from a $^{12}$C nucleus, retaining most of its energy. The back-scattered $^{12}$C introduces a large energy deposit into anodes 6 and 7 while the forward-scattered $^{20}$O provides beam-like ionization in the subsequent section of the detector. 
Two-body events are further distinguished by utilizing the left-right segmentation of the anodes \cite{Carnelli14, Johnstone21}.

\begin{figure}
\includegraphics[scale=0.40]{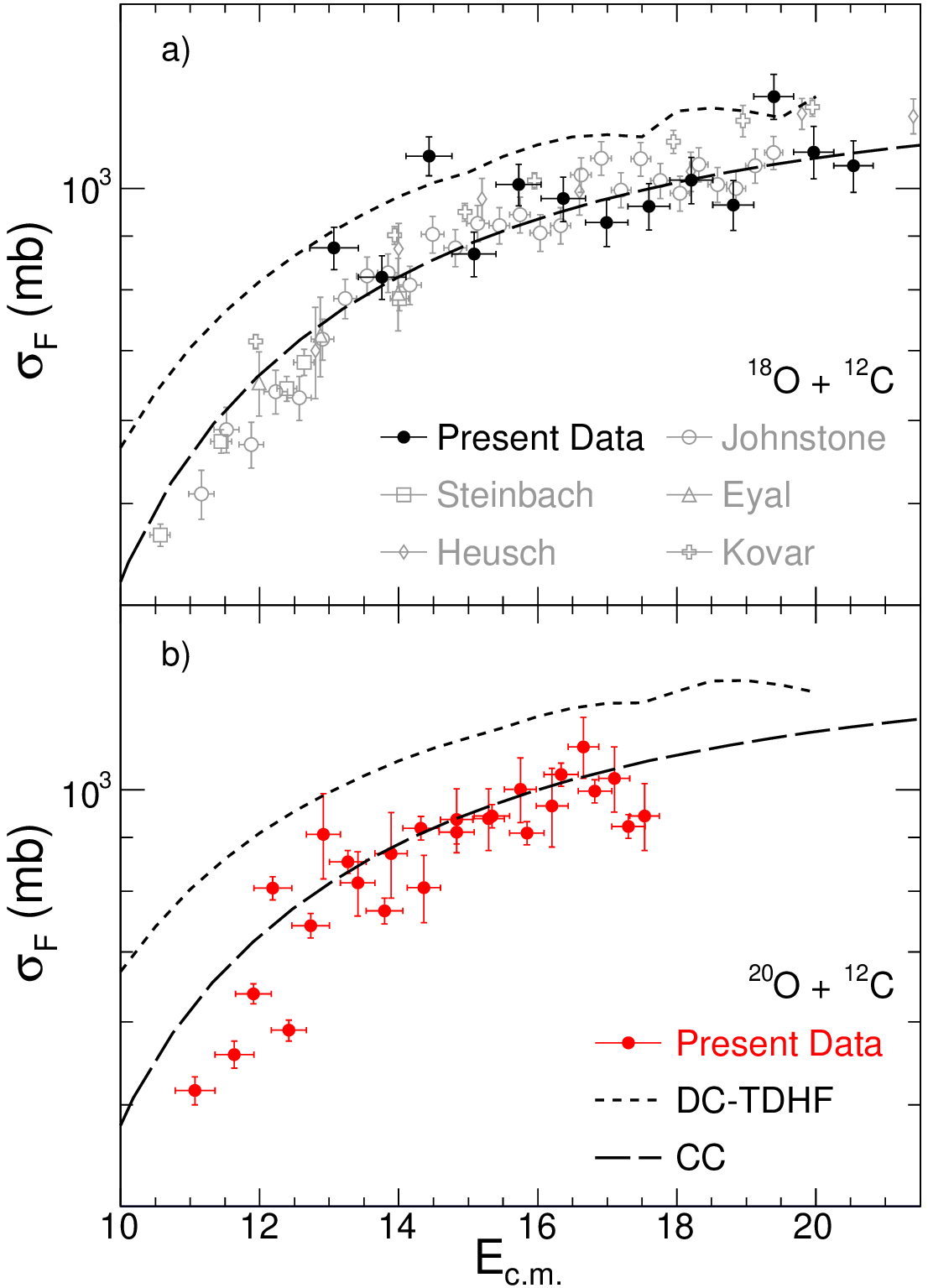}
\caption{ 
Panel a) Comparison of the excitation function for $^{18}$O + $^{12}$C measured in the present experiment (solid symbols) along with prior measurements (open symbols) \cite{Eyal76, Kovar79, Heusch82, Steinbach14a, Johnstone22}.
Panel b) Fusion excitation function for  $^{20}$O + $^{12}$C.
}
\label{fig:Exfunc}
\end{figure}

To validate the capability of the present setup to accurately measure the fusion cross-section, 
the excitation function for $^{18}$O + $^{12}$C was measured with $^{18}$O ions incident at E/A = 3.3 MeV. The resulting excitation function is displayed in Fig.~\ref{fig:Exfunc}a. Relatively good agreement of the present measurement (closed symbols) with the previously measured data is observed confirming the ability of the present experimental setup to accurately measure the fusion excitation function. When both vertical and horizontal error bars are considered only two points, at E$_{c.m.}$$\sim$13.1 and 14.4 MeV, deviate significantly from the previously published data. The deviation of these points is not presently understood.
Shown in Fig.~\ref{fig:Exfunc}b is the first measurement of the fusion excitation function for $^{20}$O + $^{12}$C in the energy interval 10.6 MeV $\leq$E$_{c.m.}$$\leq$17.5 MeV. As expected, the excitation function manifests a general decrease with decreasing energy, consistent with a barrier controlled process. A sharp drop in cross-section is observed at E$_{c.m.}$$\approx$ 13 MeV which might signal a transition in the potential associated with individual {\em{l}}-waves \cite{Esbensen12, Simenel13, deSouza23} or presence of a resonance.

\begin{figure}
\includegraphics[scale=0.35]{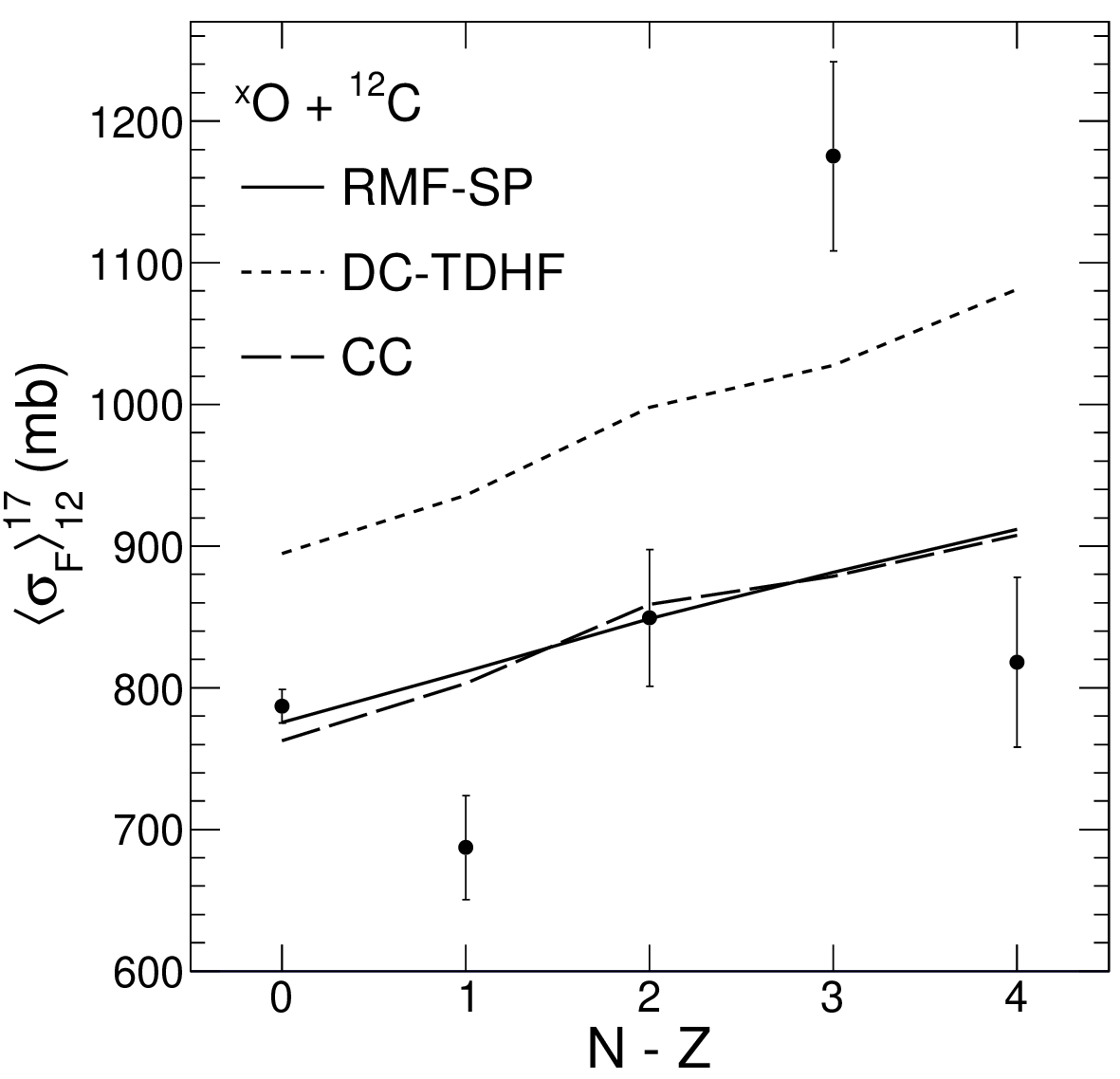}
\caption{ 
Dependence of the average cross-section on neutron excess. Error bars reflect both the statistical and systematic uncertainties in the measurements. Data are taken from: $^{16}$O \cite{Eyal76, Cujec76, Kovar79, Frawley82}; $^{17}$O \cite{Hudan23, Eyal76, Tighe93, Hertz78, Asher21}; $^{18}$O \cite{Johnstone22, Steinbach14a, Eyal76, Kovar79, Heusch82}; and $^{19}$O \cite{Hudan20}.
}
\label{fig:ave_xsect}
\end{figure}

To examine the dependence of the fusion cross-section as a function of neutron excess we calculate the average fusion cross-section in the interval 12 MeV$\leq$E$_{c.m.}$$\leq$17 MeV. 
The choice of this energy interval is dictated by the region for which the fusion excitation function is measured with the high energy bound dictated by the $^{20}$O data. For this reason, the energy interval used is slightly smaller than in prior work \cite{Hudan20, Hudan23}. In Fig.~\ref{fig:ave_xsect} the present experimental results are compared with the prior data for $^{16-19}$O + $^{12}$C \cite{Hudan20}. Surprisingly, the $^{20}$O data does not manifest the large increase previously observed for $^{19}$O but exhibits a cross-section that is essentially the same as that of  $^{18}$O. The experimental data shown indicates that based on the average fusion cross-section different categories emerge. One category is the case of the even neutron number (spin-paired) nuclei $^{16,18,20}$O which all exhibit essentially the same cross-section. The other category is that of the odd neutron number (spin-unpaired) nuclei $^{17}$O and $^{19}$O. In this latter category $^{17}$O manifests a suppression of the fusion cross-section relative to the spin-paired nuclei while $^{19}$O exhibits an enhancement. Interpretation of the data is aided by comparison with predictions of three theoretical models.

A simple approach to describe the fusion cross-section is the Sao Paulo model which involves a 
double-folding of the ground-state density distribution of the colliding nuclei \cite{Gasques04}. 
The frozen density distributions were calculated within a relativistic mean field (RMF) approach using the FSUGOLD interaction \cite{Ring96, Serot86} and the resulting average fusion cross-section is presented as the solid line (RMF-SP). 
It provides a completely static reference for the change in the fusion cross-section with increasing neutron excess by only accounting for the increased size of the colliding nuclei. To address the impact of dynamics we also performed coupled channels calculations which consider fusion dynamics through coupling to excited states due to mutual Coulomb excitation of the colliding nuclei. 
Calculations with this coupled-channels approach have been relatively successful at describing the fusion of nuclei at and near stability \cite{Montagnoli17}.
Predictions of the coupled channels model, CC, are depicted as the dashed lines in both 
Fig.~\ref{fig:Exfunc} and Fig.~\ref{fig:ave_xsect}. As evident in Fig.~\ref{fig:Exfunc}, the CC calculations are in good agreement with the experimental data for both $^{18}$O and $^{20}$O. For the lowest energies measured E$\leq$13 MeV the CC calculations slightly overpredict the measured data.

In the present CC calculations only the first exited state is included. Comparison of the cross-section with and without the first excited state led us to ignore the contribution of higher-lying states which contribute progressively less to the fusion cross-section. 
For both the RMF-SP and CC calculations the dependence of the average fusion on neutron excess is similar. To examine the role of dynamics within a different framework we also performed density constrained time-dependent Hartree-Fock (DC-TDHF) calculations. Depicted in Fig.~\ref{fig:Exfunc} as the dotted line, one observes that the DC-TDHF cross-sections are systematically higher than the experimental data although the shape of the excitation function is reasonably reproduced. 
In contrast to the RMF-SP and CC models the DC-TDHF predicts a significantly larger average fusion cross-section indicative of the difference in the initial neutron and proton density distributions and to a lesser extent the dynamics in the DC-TDHF model \cite{Hudan20}.

With the reference provided by the RMF-SP and CC calculations we note that 
the cross-section for $^{16}$O and $^{18}$O is reasonably well described while
the slight decrease experimentally observed from $^{18}$O to $^{20}$O is less than the increase predicted by  from the RMF-SP model which reflects the change in the static size. 

The fusion cross-section for $^{17}$O and $^{19}$O with an unpaired neutron deviates significantly from the RMF-SP and CC model predictions. 
It is particularly noteworthy that this deviation is in opposite directions for the two nuclei with a suppression evident
for $^{17}$O and an enhancement observed for $^{19}$O, even though the one neutron separation energies of 4143 keV and 3956 keV 
for $^{17}$O and $^{19}$O respectively are essentially the same. The DC-TDHF model, which includes dynamics, provides a significantly larger cross-section than the CC calculations. While it is in better agreement with the $^{19}$O cross-section it significantly over-predicts both the neutron-paired cases of $^{16,18,20}$O as well as unpaired neutron case of $^{17}$O. These results for both the spin-paired and spin-unpaired cases suggest that for this isotopic chain the near-barrier fusion cross-sections cannot be understood within the framework of a simple barrier which systematically evolves.

Given the relatively strong binding of the neutrons in  $^{20}$O it is unlikely that neutron breakup is on average responsible for this near constancy of the average cross-section for $^{20}$O as compared to $^{18}$O. Rather, it is likely that the barrier/barrier distributions for the different nuclei do not follow a simple systematic trend. If the barrier/barrier distributions change significantly from one nucleus to another, reflecting the influence of initial structure and pairing on fusion, then calculation of an integrated cross-section over a common interval could introduce the observed behavior. Direct extraction of barrier distributions, possibly through the double derivative approach \cite{DasGupta98}, requires higher resolution data than is presently available, motivating future experiments.

Use of an active target technique allowed an effective first measurement of the fusion excitation function for $^{20}$O + $^{12}$C at near-barrier energies. A large increase of the average fusion cross-section previously observed for $^{19}$O is not observed for $^{20}$O.  Rather, over the interval 12 MeV $\leq$E$_{c.m.}$$\leq$17 MeV, the average fusion cross-section, $<$$\sigma_F$$>$$_{12}^{17}$, for $^{20}$O is slightly decreased as compared to $^{18}$O. In marked contrast to the even neutron number (N) cases, nuclei with an odd number of neutrons exhibit significantly different behavior.  The deviation of the odd-N as compared to even-N nuclei could reflect differences in pairing/pairing dynamics or the initial shell structure that persist through fusion. Both static and dynamic theoretical models are unable to describe the behavior of the average fusion cross-section with neutron-excess for the entire isotopic chain. Future high-resolution measurements of the fusion excitation function for this isotopic chain with both paired and unpaired neutrons are critical in understanding fusion for neutron-rich light nuclei and investigating the role of initial structure and pairing on fusion.

\begin{acknowledgments}
We acknowledge the high quality beam and experimental support provided by the technical and scientific staff at the Grand Acc\'{e}l\'{e}rateur National d'Ions Lourds (GANIL) that made this experiment possible. In particular we appreciate the assistance of D. Allal, B. Jacquot, and D. Gruyer. Scientific discussions with N. Alahari are gratefully acknowledged. 
We are thankful for the high-quality services of the Mechanical Instrument Services and Electronic Instrument Services facilities at Indiana University.
This work was supported by the U.S. Department of Energy Office of Science under Grant No. 
DE-FG02-88ER-40404 and Indiana University. The research leading to these results has received funding from the European Union's HORIZON EUROPE Program under grant agreement n°101057511. R. deSouza gratefully acknowledges the support of the GANIL Visiting Scientist Program. Kyle Brown acknowledges support from Michigan State University.

\end{acknowledgments}


\begin{thebibliography}{43}%
\makeatletter
\providecommand \@ifxundefined [1]{%
 \@ifx{#1\undefined}
}%
\providecommand \@ifnum [1]{%
 \ifnum #1\expandafter \@firstoftwo
 \else \expandafter \@secondoftwo
 \fi
}%
\providecommand \@ifx [1]{%
 \ifx #1\expandafter \@firstoftwo
 \else \expandafter \@secondoftwo
 \fi
}%
\providecommand \natexlab [1]{#1}%
\providecommand \enquote  [1]{``#1''}%
\providecommand \bibnamefont  [1]{#1}%
\providecommand \bibfnamefont [1]{#1}%
\providecommand \citenamefont [1]{#1}%
\providecommand \href@noop [0]{\@secondoftwo}%
\providecommand \href [0]{\begingroup \@sanitize@url \@href}%
\providecommand \@href[1]{\@@startlink{#1}\@@href}%
\providecommand \@@href[1]{\endgroup#1\@@endlink}%
\providecommand \@sanitize@url [0]{\catcode `\\12\catcode `\$12\catcode
  `\&12\catcode `\#12\catcode `\^12\catcode `\_12\catcode `\%12\relax}%
\providecommand \@@startlink[1]{}%
\providecommand \@@endlink[0]{}%
\providecommand \url  [0]{\begingroup\@sanitize@url \@url }%
\providecommand \@url [1]{\endgroup\@href {#1}{\urlprefix }}%
\providecommand \urlprefix  [0]{URL }%
\providecommand \Eprint [0]{\href }%
\providecommand \doibase [0]{http://dx.doi.org/}%
\providecommand \selectlanguage [0]{\@gobble}%
\providecommand \bibinfo  [0]{\@secondoftwo}%
\providecommand \bibfield  [0]{\@secondoftwo}%
\providecommand \translation [1]{[#1]}%
\providecommand \BibitemOpen [0]{}%
\providecommand \bibitemStop [0]{}%
\providecommand \bibitemNoStop [0]{.\EOS\space}%
\providecommand \EOS [0]{\spacefactor3000\relax}%
\providecommand \BibitemShut  [1]{\csname bibitem#1\endcsname}%
\let\auto@bib@innerbib\@empty
\bibitem [{\citenamefont {Nicholl}\ \emph {et~al.}(2017)\citenamefont {Nicholl}
  \emph {et~al.}}]{Nicholl17}%
  \BibitemOpen
  \bibfield  {author} {\bibinfo {author} {\bibfnamefont {M.}~\bibnamefont
  {Nicholl}} \emph {et~al.},\ }\href {\doibase 10.3847/2041-8213/aa9029}
  {\bibfield  {journal} {\bibinfo  {journal} {Astrophys. J. Lett.}\ }\textbf
  {\bibinfo {volume} {848}},\ \bibinfo {pages} {L18} (\bibinfo {year}
  {2017})}\BibitemShut {NoStop}%
\bibitem [{\citenamefont {Casal}\ \emph {et~al.}(2020)\citenamefont {Casal},
  \citenamefont {Singh}, \citenamefont {Fortunato}, \citenamefont {Horiuchi},\
  and\ \citenamefont {Vitturi}}]{Casal20}%
  \BibitemOpen
  \bibfield  {author} {\bibinfo {author} {\bibfnamefont {J.}~\bibnamefont
  {Casal}}, \bibinfo {author} {\bibfnamefont {J.}~\bibnamefont {Singh}},
  \bibinfo {author} {\bibfnamefont {L.}~\bibnamefont {Fortunato}}, \bibinfo
  {author} {\bibfnamefont {W.}~\bibnamefont {Horiuchi}}, \ and\ \bibinfo
  {author} {\bibfnamefont {A.}~\bibnamefont {Vitturi}},\ }\href {\doibase
  10.1103/PhysRevC.102.064627} {\bibfield  {journal} {\bibinfo  {journal}
  {Phys. Rev. C}\ }\textbf {\bibinfo {volume} {102}},\ \bibinfo {pages}
  {064627} (\bibinfo {year} {2020})}\BibitemShut {NoStop}%
\bibitem [{\citenamefont {Michel}\ \emph {et~al.}(2008)\citenamefont {Michel},
  \citenamefont {Nazarewicz}, \citenamefont {Płoszajczak},\ and\ \citenamefont
  {Vertse}}]{Michel08}%
  \BibitemOpen
  \bibfield  {author} {\bibinfo {author} {\bibfnamefont {N.}~\bibnamefont
  {Michel}}, \bibinfo {author} {\bibfnamefont {W.}~\bibnamefont {Nazarewicz}},
  \bibinfo {author} {\bibfnamefont {M.}~\bibnamefont {Płoszajczak}}, \ and\
  \bibinfo {author} {\bibfnamefont {T.}~\bibnamefont {Vertse}},\ }\href
  {\doibase 10.1088/0954-3899/36/1/013101} {\bibfield  {journal} {\bibinfo
  {journal} {J. Phys. G}\ }\textbf {\bibinfo {volume} {36}},\ \bibinfo {pages}
  {013101} (\bibinfo {year} {2008})}\BibitemShut {NoStop}%
\bibitem [{\citenamefont {Nakatsuka}\ \emph {et~al.}(2017)\citenamefont
  {Nakatsuka} \emph {et~al.}}]{Nakatsuka17}%
  \BibitemOpen
  \bibfield  {author} {\bibinfo {author} {\bibfnamefont {N.}~\bibnamefont
  {Nakatsuka}} \emph {et~al.},\ }\href {\doibase
  10.1016/j.physletb.2017.03.017} {\bibfield  {journal} {\bibinfo  {journal}
  {Phys. Lett. B}\ }\textbf {\bibinfo {volume} {768}},\ \bibinfo {pages} {387}
  (\bibinfo {year} {2017})}\BibitemShut {NoStop}%
\bibitem [{\citenamefont {Magierski}\ \emph {et~al.}(2017)\citenamefont
  {Magierski}, \citenamefont {Sekizawa},\ and\ \citenamefont
  {Wlazłowski}}]{Magierski17}%
  \BibitemOpen
  \bibfield  {author} {\bibinfo {author} {\bibfnamefont {P.}~\bibnamefont
  {Magierski}}, \bibinfo {author} {\bibfnamefont {K.}~\bibnamefont {Sekizawa}},
  \ and\ \bibinfo {author} {\bibfnamefont {G.}~\bibnamefont {Wlazłowski}},\
  }\href {\doibase 10.1103/PhysRevLett.119.042501} {\bibfield  {journal}
  {\bibinfo  {journal} {Phys. Rev. Lett.}\ }\textbf {\bibinfo {volume} {119}},\
  \bibinfo {pages} {042501} (\bibinfo {year} {2017})}\BibitemShut {NoStop}%
\bibitem [{\citenamefont {Scamps}\ and\ \citenamefont
  {Hashimoto}(2019)}]{Scamps19}%
  \BibitemOpen
  \bibfield  {author} {\bibinfo {author} {\bibfnamefont {G.}~\bibnamefont
  {Scamps}}\ and\ \bibinfo {author} {\bibfnamefont {Y.}~\bibnamefont
  {Hashimoto}},\ }\href {\doibase 10.1103/PhysRevC.100.024623} {\bibfield
  {journal} {\bibinfo  {journal} {Phys. Rev. C}\ }\textbf {\bibinfo {volume}
  {100}},\ \bibinfo {pages} {024623} (\bibinfo {year} {2019})}\BibitemShut
  {NoStop}%
\bibitem [{\citenamefont {Rehm}()}]{Hagino21}%
  \BibitemOpen
  \bibfield  {author} {\bibinfo {author} {\bibfnamefont {K.}~\bibnamefont
  {Rehm}},\ }\href@noop {} {}\bibinfo {howpublished} {Private
  Communication}\BibitemShut {NoStop}%
\bibitem [{\citenamefont {Cortes}\ \emph {et~al.}(2020)\citenamefont {Cortes},
  \citenamefont {Rangel}, \citenamefont {Ferreira}, \citenamefont {Lubian},\
  and\ \citenamefont {Canto}}]{Cortes20}%
  \BibitemOpen
  \bibfield  {author} {\bibinfo {author} {\bibfnamefont {M.~R.}\ \bibnamefont
  {Cortes}}, \bibinfo {author} {\bibfnamefont {J.}~\bibnamefont {Rangel}},
  \bibinfo {author} {\bibfnamefont {J.~L.}\ \bibnamefont {Ferreira}}, \bibinfo
  {author} {\bibfnamefont {J.}~\bibnamefont {Lubian}}, \ and\ \bibinfo {author}
  {\bibfnamefont {L.~F.}\ \bibnamefont {Canto}},\ }\href {\doibase
  doi.org/10.1103/PhysRevC.102.064628} {\bibfield  {journal} {\bibinfo
  {journal} {Phys. Rev. C}\ }\textbf {\bibinfo {volume} {102}},\ \bibinfo
  {pages} {064628} (\bibinfo {year} {2020})}\BibitemShut {NoStop}%
\bibitem [{\citenamefont {FRIB}()}]{FRIB}%
  \BibitemOpen
  \bibfield  {author} {\bibinfo {author} {\bibnamefont {FRIB}},\ }\href
  {http://frib.msu.edu} {}\bibinfo {note} {{F}acility for {R}are {I}sotope
  {B}eams, {M}ichigan {S}tate {U}niversity, USA}\BibitemShut {NoStop}%
\bibitem [{\citenamefont {RIKEN}()}]{RIKEN}%
  \BibitemOpen
  \bibfield  {author} {\bibinfo {author} {\bibnamefont {RIKEN}},\ }\href
  {http://www.riken.jp/} {}\bibinfo {note} {{N}ishina {C}enter for
  {A}ccelerator-{B}ased {S}cience, {J}apan}\BibitemShut {NoStop}%
\bibitem [{\citenamefont {GANIL}()}]{GANIL}%
  \BibitemOpen
  \bibfield  {author} {\bibinfo {author} {\bibnamefont {GANIL}},\ }\href
  {http://www.ganil-spiral2.eu/} {}\bibinfo {note} {{G}rand {A}ccel\'erateur
  {N}ational d'{I}ons {L}ourds, {C}aen, {F}rance}\BibitemShut {NoStop}%
\bibitem [{\citenamefont {Gasques}\ \emph {et~al.}(2007)\citenamefont
  {Gasques}, \citenamefont {Afanasjev}, \citenamefont {Beard}, \citenamefont
  {Lubian}, \citenamefont {Neff}, \citenamefont {Wiescher},\ and\ \citenamefont
  {Yakolev}}]{Gasques07}%
  \BibitemOpen
  \bibfield  {author} {\bibinfo {author} {\bibfnamefont {L.~R.}\ \bibnamefont
  {Gasques}}, \bibinfo {author} {\bibfnamefont {A.}~\bibnamefont {Afanasjev}},
  \bibinfo {author} {\bibfnamefont {M.}~\bibnamefont {Beard}}, \bibinfo
  {author} {\bibfnamefont {M.~J.}\ \bibnamefont {Lubian}}, \bibinfo {author}
  {\bibfnamefont {T.}~\bibnamefont {Neff}}, \bibinfo {author} {\bibfnamefont
  {M.}~\bibnamefont {Wiescher}}, \ and\ \bibinfo {author} {\bibfnamefont
  {D.}~\bibnamefont {Yakolev}},\ }\href {\doibase 10.1103/PhysRevC.76.045802}
  {\bibfield  {journal} {\bibinfo  {journal} {Phys. Rev. C}\ }\textbf {\bibinfo
  {volume} {76}},\ \bibinfo {pages} {045802} (\bibinfo {year}
  {2007})}\BibitemShut {NoStop}%
\bibitem [{\citenamefont {Steinbach}\ \emph {et~al.}(2014)\citenamefont
  {Steinbach}, \citenamefont {Vadas}, \citenamefont {Schmidt}, \citenamefont
  {Haycraft}, \citenamefont {Hudan}, \citenamefont {deSouza}, \citenamefont
  {Baby}, \citenamefont {Kuvin}, \citenamefont {Wiedenh\"over}, \citenamefont
  {Umar},\ and\ \citenamefont {Oberacker}}]{Steinbach14a}%
  \BibitemOpen
  \bibfield  {author} {\bibinfo {author} {\bibfnamefont {T.~K.}\ \bibnamefont
  {Steinbach}}, \bibinfo {author} {\bibfnamefont {J.}~\bibnamefont {Vadas}},
  \bibinfo {author} {\bibfnamefont {J.}~\bibnamefont {Schmidt}}, \bibinfo
  {author} {\bibfnamefont {C.}~\bibnamefont {Haycraft}}, \bibinfo {author}
  {\bibfnamefont {S.}~\bibnamefont {Hudan}}, \bibinfo {author} {\bibfnamefont
  {R.~T.}\ \bibnamefont {deSouza}}, \bibinfo {author} {\bibfnamefont {L.~T.}\
  \bibnamefont {Baby}}, \bibinfo {author} {\bibfnamefont {S.~A.}\ \bibnamefont
  {Kuvin}}, \bibinfo {author} {\bibfnamefont {I.}~\bibnamefont
  {Wiedenh\"over}}, \bibinfo {author} {\bibfnamefont {A.~S.}\ \bibnamefont
  {Umar}}, \ and\ \bibinfo {author} {\bibfnamefont {V.~E.}\ \bibnamefont
  {Oberacker}},\ }\href {\doibase 10.1103/PhysRevC.90.041603} {\bibfield
  {journal} {\bibinfo  {journal} {Phys. Rev. C}\ }\textbf {\bibinfo {volume}
  {90}},\ \bibinfo {pages} {041603(R)} (\bibinfo {year} {2014})}\BibitemShut
  {NoStop}%
\bibitem [{\citenamefont {{Varinderjit Singh}}\ \emph
  {et~al.}(2017)\citenamefont {{Varinderjit Singh}}, \citenamefont {Vadas},
  \citenamefont {Steinbach}, \citenamefont {Wiggins}, \citenamefont {Hudan},
  \citenamefont {{deSouza}}, \citenamefont {{Zidu Lin}}, \citenamefont
  {Horowitz}, \citenamefont {Baby}, \citenamefont {Kuvin}, \citenamefont
  {{Vandana Tripathi}}, \citenamefont {Wiedenh\"over},\ and\ \citenamefont
  {Umar}}]{Singh17}%
  \BibitemOpen
  \bibfield  {author} {\bibinfo {author} {\bibnamefont {{Varinderjit Singh}}},
  \bibinfo {author} {\bibfnamefont {J.}~\bibnamefont {Vadas}}, \bibinfo
  {author} {\bibfnamefont {T.~K.}\ \bibnamefont {Steinbach}}, \bibinfo {author}
  {\bibfnamefont {B.~B.}\ \bibnamefont {Wiggins}}, \bibinfo {author}
  {\bibfnamefont {S.}~\bibnamefont {Hudan}}, \bibinfo {author} {\bibfnamefont
  {R.~T.}\ \bibnamefont {{deSouza}}}, \bibinfo {author} {\bibnamefont {{Zidu
  Lin}}}, \bibinfo {author} {\bibfnamefont {C.~J.}\ \bibnamefont {Horowitz}},
  \bibinfo {author} {\bibfnamefont {L.~T.}\ \bibnamefont {Baby}}, \bibinfo
  {author} {\bibfnamefont {S.~A.}\ \bibnamefont {Kuvin}}, \bibinfo {author}
  {\bibnamefont {{Vandana Tripathi}}}, \bibinfo {author} {\bibfnamefont
  {I.}~\bibnamefont {Wiedenh\"over}}, \ and\ \bibinfo {author} {\bibfnamefont
  {A.~S.}\ \bibnamefont {Umar}},\ }\href {\doibase
  10.1016/j.physletb.2016.12.017} {\bibfield  {journal} {\bibinfo  {journal}
  {Phys. Lett. B}\ }\textbf {\bibinfo {volume} {765}},\ \bibinfo {pages} {99}
  (\bibinfo {year} {2017})}\BibitemShut {NoStop}%
\bibitem [{\citenamefont {Vadas}\ \emph {et~al.}(2018)\citenamefont {Vadas},
  \citenamefont {Singh}, \citenamefont {Wiggins}, \citenamefont {Huston},
  \citenamefont {Hudan}, \citenamefont {deSouza}, \citenamefont {Lin},
  \citenamefont {Horowitz}, \citenamefont {Chbihi}, \citenamefont {Ackermann},
  \citenamefont {Famiano},\ and\ \citenamefont {Brown}}]{Vadas18}%
  \BibitemOpen
  \bibfield  {author} {\bibinfo {author} {\bibfnamefont {J.}~\bibnamefont
  {Vadas}}, \bibinfo {author} {\bibfnamefont {V.}~\bibnamefont {Singh}},
  \bibinfo {author} {\bibfnamefont {B.~B.}\ \bibnamefont {Wiggins}}, \bibinfo
  {author} {\bibfnamefont {J.}~\bibnamefont {Huston}}, \bibinfo {author}
  {\bibfnamefont {S.}~\bibnamefont {Hudan}}, \bibinfo {author} {\bibfnamefont
  {R.~T.}\ \bibnamefont {deSouza}}, \bibinfo {author} {\bibfnamefont
  {Z.}~\bibnamefont {Lin}}, \bibinfo {author} {\bibfnamefont {C.~J.}\
  \bibnamefont {Horowitz}}, \bibinfo {author} {\bibfnamefont {A.}~\bibnamefont
  {Chbihi}}, \bibinfo {author} {\bibfnamefont {D.}~\bibnamefont {Ackermann}},
  \bibinfo {author} {\bibfnamefont {M.}~\bibnamefont {Famiano}}, \ and\
  \bibinfo {author} {\bibfnamefont {K.~W.}\ \bibnamefont {Brown}},\ }\href
  {\doibase 10.1103/PhysRevC.97.031601} {\bibfield  {journal} {\bibinfo
  {journal} {Phys. Rev. C}\ }\textbf {\bibinfo {volume} {97}},\ \bibinfo
  {pages} {031601(R)} (\bibinfo {year} {2018})}\BibitemShut {NoStop}%
\bibitem [{\citenamefont {Hudan}\ \emph {et~al.}(2020)\citenamefont {Hudan},
  \citenamefont {{deSouza}}, \citenamefont {{A.~S. Umar}}, \citenamefont {{Zidu
  Lin}},\ and\ \citenamefont {Horowitz}}]{Hudan20}%
  \BibitemOpen
  \bibfield  {author} {\bibinfo {author} {\bibfnamefont {S.}~\bibnamefont
  {Hudan}}, \bibinfo {author} {\bibfnamefont {R.~T.}\ \bibnamefont
  {{deSouza}}}, \bibinfo {author} {\bibnamefont {{A.~S. Umar}}}, \bibinfo
  {author} {\bibnamefont {{Zidu Lin}}}, \ and\ \bibinfo {author} {\bibfnamefont
  {C.~J.}\ \bibnamefont {Horowitz}},\ }\href {\doibase
  10.1103/PhysRevC.101.061601} {\bibfield  {journal} {\bibinfo  {journal}
  {Phys. Rev. C}\ }\textbf {\bibinfo {volume} {101}},\ \bibinfo {pages}
  {061601} (\bibinfo {year} {2020})}\BibitemShut {NoStop}%
\bibitem [{\citenamefont {Tanihata}\ \emph
  {et~al.}(1985{\natexlab{a}})\citenamefont {Tanihata} \emph
  {et~al.}}]{Tanihata85a}%
  \BibitemOpen
  \bibfield  {author} {\bibinfo {author} {\bibfnamefont {I.}~\bibnamefont
  {Tanihata}} \emph {et~al.},\ }\href {\doibase 10.1016/0370-2693(85)90005-X}
  {\bibfield  {journal} {\bibinfo  {journal} {Phys. Lett. B}\ }\textbf
  {\bibinfo {volume} {160}},\ \bibinfo {pages} {380} (\bibinfo {year}
  {1985}{\natexlab{a}})}\BibitemShut {NoStop}%
\bibitem [{\citenamefont {Tanihata}\ \emph
  {et~al.}(1985{\natexlab{b}})\citenamefont {Tanihata} \emph
  {et~al.}}]{Tanihata85b}%
  \BibitemOpen
  \bibfield  {author} {\bibinfo {author} {\bibfnamefont {I.}~\bibnamefont
  {Tanihata}} \emph {et~al.},\ }\href {\doibase 10.1103/physrevlett.55.2676}
  {\bibfield  {journal} {\bibinfo  {journal} {Phys. Rev. Lett.}\ }\textbf
  {\bibinfo {volume} {55}},\ \bibinfo {pages} {2676} (\bibinfo {year}
  {1985}{\natexlab{b}})}\BibitemShut {NoStop}%
\bibitem [{\citenamefont {Carnelli}\ \emph {et~al.}(2015)\citenamefont
  {Carnelli} \emph {et~al.}}]{Carnelli15}%
  \BibitemOpen
  \bibfield  {author} {\bibinfo {author} {\bibfnamefont {P.~F.~F.}\
  \bibnamefont {Carnelli}} \emph {et~al.},\ }\href {\doibase
  10.1016/j.nima.2015.07.030} {\bibfield  {journal} {\bibinfo  {journal} {Nucl.
  Instr. Meth. A}\ }\textbf {\bibinfo {volume} {799}},\ \bibinfo {pages} {197}
  (\bibinfo {year} {2015})}\BibitemShut {NoStop}%
\bibitem [{\citenamefont {Almaraz-Calderon}\ \emph {et~al.}(2015)\citenamefont
  {Almaraz-Calderon} \emph {et~al.}}]{Calderon15}%
  \BibitemOpen
  \bibfield  {author} {\bibinfo {author} {\bibfnamefont {S.}~\bibnamefont
  {Almaraz-Calderon}} \emph {et~al.},\ }\href {\doibase
  10.1051/epjconf/20159601001} {\bibfield  {journal} {\bibinfo  {journal} {EPJ
  Web of Conferences}\ }\textbf {\bibinfo {volume} {96}},\ \bibinfo {pages}
  {01001} (\bibinfo {year} {2015})}\BibitemShut {NoStop}%
\bibitem [{\citenamefont {deSouza}\ \emph {et~al.}(2021)\citenamefont {deSouza}
  \emph {et~al.}}]{deSouza21}%
  \BibitemOpen
  \bibfield  {author} {\bibinfo {author} {\bibfnamefont {R.~T.}\ \bibnamefont
  {deSouza}} \emph {et~al.},\ }\href {\doibase 10.1016/j.physletb.2021.136115}
  {\bibfield  {journal} {\bibinfo  {journal} {Physs. Lett. B}\ }\textbf
  {\bibinfo {volume} {814}},\ \bibinfo {pages} {136115} (\bibinfo {year}
  {2021})}\BibitemShut {NoStop}%
\bibitem [{\citenamefont {Shapira}\ \emph {et~al.}(2000)\citenamefont
  {Shapira}, \citenamefont {Lewis},\ and\ \citenamefont {Hulett}}]{Shapira00}%
  \BibitemOpen
  \bibfield  {author} {\bibinfo {author} {\bibfnamefont {D.}~\bibnamefont
  {Shapira}}, \bibinfo {author} {\bibfnamefont {T.~A.}\ \bibnamefont {Lewis}},
  \ and\ \bibinfo {author} {\bibfnamefont {L.~D.}\ \bibnamefont {Hulett}},\
  }\href {\doibase 10.1016/S0168-9002(00)00499-X} {\bibfield  {journal}
  {\bibinfo  {journal} {Nucl. Instr. Meth. A}\ }\textbf {\bibinfo {volume}
  {454}},\ \bibinfo {pages} {409} (\bibinfo {year} {2000})}\BibitemShut
  {NoStop}%
\bibitem [{\citenamefont {Carnelli}\ \emph {et~al.}(2014)\citenamefont
  {Carnelli} \emph {et~al.}}]{Carnelli14}%
  \BibitemOpen
  \bibfield  {author} {\bibinfo {author} {\bibfnamefont {P.~F.~F.}\
  \bibnamefont {Carnelli}} \emph {et~al.},\ }\href {\doibase
  10.1103/PhysRevLett.112.192701} {\bibfield  {journal} {\bibinfo  {journal}
  {Phys. Rev. Lett.}\ }\textbf {\bibinfo {volume} {112}},\ \bibinfo {pages}
  {192701} (\bibinfo {year} {2014})}\BibitemShut {NoStop}%
\bibitem [{\citenamefont {Johnstone}\ \emph
  {et~al.}(2021{\natexlab{a}})\citenamefont {Johnstone} \emph
  {et~al.}}]{Johnstone21}%
  \BibitemOpen
  \bibfield  {author} {\bibinfo {author} {\bibfnamefont {J.}~\bibnamefont
  {Johnstone}} \emph {et~al.},\ }\href {\doibase 10.1016/j.nima.2021.165697}
  {\bibfield  {journal} {\bibinfo  {journal} {Nucl. Instr. Meth. A}\ }\textbf
  {\bibinfo {volume} {1014}},\ \bibinfo {pages} {166697} (\bibinfo {year}
  {2021}{\natexlab{a}})}\BibitemShut {NoStop}%
\bibitem [{\citenamefont {Foreman}\ \emph {et~al.}(1981)\citenamefont
  {Foreman}, \citenamefont {Kleban}, \citenamefont {Schmidt},\ and\
  \citenamefont {Davis}}]{Foreman81}%
  \BibitemOpen
  \bibfield  {author} {\bibinfo {author} {\bibfnamefont {L.}~\bibnamefont
  {Foreman}}, \bibinfo {author} {\bibfnamefont {P.}~\bibnamefont {Kleban}},
  \bibinfo {author} {\bibfnamefont {L.}~\bibnamefont {Schmidt}}, \ and\
  \bibinfo {author} {\bibfnamefont {H.}~\bibnamefont {Davis}},\ }\href
  {\doibase 10.1103/PhysRevA.23.1553} {\bibfield  {journal} {\bibinfo
  {journal} {Phys. Rev. A}\ }\textbf {\bibinfo {volume} {23}},\ \bibinfo
  {pages} {1553} (\bibinfo {year} {1981})}\BibitemShut {NoStop}%
\bibitem [{\citenamefont {Johnstone}\ \emph
  {et~al.}(2021{\natexlab{b}})\citenamefont {Johnstone} \emph
  {et~al.}}]{Johnstone22}%
  \BibitemOpen
  \bibfield  {author} {\bibinfo {author} {\bibfnamefont {J.}~\bibnamefont
  {Johnstone}} \emph {et~al.},\ }\href {\doibase 10.1016/j.nima.2022.166212}
  {\bibfield  {journal} {\bibinfo  {journal} {Nucl. Instr. Meth. A}\ }\textbf
  {\bibinfo {volume} {1025}},\ \bibinfo {pages} {166212} (\bibinfo {year}
  {2021}{\natexlab{b}})}\BibitemShut {NoStop}%
\bibitem [{\citenamefont {Eyal}\ \emph {et~al.}(1976)\citenamefont {Eyal},
  \citenamefont {Beckerman}, \citenamefont {Chechik}, \citenamefont
  {Fraenkel},\ and\ \citenamefont {Stocker}}]{Eyal76}%
  \BibitemOpen
  \bibfield  {author} {\bibinfo {author} {\bibfnamefont {Y.}~\bibnamefont
  {Eyal}}, \bibinfo {author} {\bibfnamefont {M.}~\bibnamefont {Beckerman}},
  \bibinfo {author} {\bibfnamefont {R.}~\bibnamefont {Chechik}}, \bibinfo
  {author} {\bibfnamefont {Z.}~\bibnamefont {Fraenkel}}, \ and\ \bibinfo
  {author} {\bibfnamefont {H.}~\bibnamefont {Stocker}},\ }\href {\doibase
  10.1103/PhysRevC.13.1527} {\bibfield  {journal} {\bibinfo  {journal} {Phys.
  Rev. C}\ }\textbf {\bibinfo {volume} {13}},\ \bibinfo {pages} {1527}
  (\bibinfo {year} {1976})}\BibitemShut {NoStop}%
\bibitem [{\citenamefont {Kovar}\ \emph {et~al.}(1979)\citenamefont {Kovar}
  \emph {et~al.}}]{Kovar79}%
  \BibitemOpen
  \bibfield  {author} {\bibinfo {author} {\bibfnamefont {D.~G.}\ \bibnamefont
  {Kovar}} \emph {et~al.},\ }\href {\doibase 10.1103/PhysRevC.20.1305}
  {\bibfield  {journal} {\bibinfo  {journal} {Phys. Rev. C}\ }\textbf {\bibinfo
  {volume} {20}},\ \bibinfo {pages} {1305} (\bibinfo {year}
  {1979})}\BibitemShut {NoStop}%
\bibitem [{\citenamefont {Heusch}\ \emph {et~al.}(1982)\citenamefont {Heusch},
  \citenamefont {Beck}, \citenamefont {Coffin}, \citenamefont {Engelstein},
  \citenamefont {Freeman}, \citenamefont {Guillaume}, \citenamefont {Haas},\
  and\ \citenamefont {Wagner}}]{Heusch82}%
  \BibitemOpen
  \bibfield  {author} {\bibinfo {author} {\bibfnamefont {B.}~\bibnamefont
  {Heusch}}, \bibinfo {author} {\bibfnamefont {C.}~\bibnamefont {Beck}},
  \bibinfo {author} {\bibfnamefont {J.~P.}\ \bibnamefont {Coffin}}, \bibinfo
  {author} {\bibfnamefont {P.}~\bibnamefont {Engelstein}}, \bibinfo {author}
  {\bibfnamefont {R.~M.}\ \bibnamefont {Freeman}}, \bibinfo {author}
  {\bibfnamefont {G.}~\bibnamefont {Guillaume}}, \bibinfo {author}
  {\bibfnamefont {F.}~\bibnamefont {Haas}}, \ and\ \bibinfo {author}
  {\bibfnamefont {P.}~\bibnamefont {Wagner}},\ }\href {\doibase
  10.1103/PhysRevC.26.542} {\bibfield  {journal} {\bibinfo  {journal} {Phys.
  Rev. C}\ }\textbf {\bibinfo {volume} {26}},\ \bibinfo {pages} {542} (\bibinfo
  {year} {1982})}\BibitemShut {NoStop}%
\bibitem [{\citenamefont {Esbensen}(2012)}]{Esbensen12}%
  \BibitemOpen
  \bibfield  {author} {\bibinfo {author} {\bibfnamefont {H.}~\bibnamefont
  {Esbensen}},\ }\href {\doibase 10.1103/PhysRevC.85.064611} {\bibfield
  {journal} {\bibinfo  {journal} {Phys. Rev. C}\ }\textbf {\bibinfo {volume}
  {85}},\ \bibinfo {pages} {064611} (\bibinfo {year} {2012})}\BibitemShut
  {NoStop}%
\bibitem [{\citenamefont {Simenel}\ \emph {et~al.}(2013)\citenamefont
  {Simenel}, \citenamefont {Keser}, \citenamefont {Umar},\ and\ \citenamefont
  {Oberacker}}]{Simenel13}%
  \BibitemOpen
  \bibfield  {author} {\bibinfo {author} {\bibfnamefont {C.}~\bibnamefont
  {Simenel}}, \bibinfo {author} {\bibfnamefont {R.}~\bibnamefont {Keser}},
  \bibinfo {author} {\bibfnamefont {A.}~\bibnamefont {Umar}}, \ and\ \bibinfo
  {author} {\bibfnamefont {V.}~\bibnamefont {Oberacker}},\ }\href {\doibase
  10.1103/PhysRevC.88.024617} {\bibfield  {journal} {\bibinfo  {journal} {Phys.
  Rev. C}\ }\textbf {\bibinfo {volume} {88}},\ \bibinfo {pages} {024617}
  (\bibinfo {year} {2013})}\BibitemShut {NoStop}%
\bibitem [{\citenamefont {deSouza}\ \emph {et~al.}(2023)\citenamefont {deSouza}
  \emph {et~al.}}]{deSouza23}%
  \BibitemOpen
  \bibfield  {author} {\bibinfo {author} {\bibfnamefont {R.~T.}\ \bibnamefont
  {deSouza}} \emph {et~al.},\ }\href@noop {} {\bibfield  {journal} {\bibinfo
  {journal} {Phys. Rev. Lett.}\ ,\ \bibinfo {pages} {(submitted)}} (\bibinfo
  {year} {2023})}\BibitemShut {NoStop}%
\bibitem [{\citenamefont {Cujec}\ and\ \citenamefont {Barnes}(1976)}]{Cujec76}%
  \BibitemOpen
  \bibfield  {author} {\bibinfo {author} {\bibfnamefont {B.}~\bibnamefont
  {Cujec}}\ and\ \bibinfo {author} {\bibfnamefont {C.~A.}\ \bibnamefont
  {Barnes}},\ }\href {\doibase 10.1016/0375-9474(76)90370-5} {\bibfield
  {journal} {\bibinfo  {journal} {Nucl. Phys. A}\ }\textbf {\bibinfo {volume}
  {266}},\ \bibinfo {pages} {461} (\bibinfo {year} {1976})}\BibitemShut
  {NoStop}%
\bibitem [{\citenamefont {Frawley}\ \emph {et~al.}(1982)\citenamefont
  {Frawley}, \citenamefont {Fletcher},\ and\ \citenamefont
  {Dennis}}]{Frawley82}%
  \BibitemOpen
  \bibfield  {author} {\bibinfo {author} {\bibfnamefont {A.~D.}\ \bibnamefont
  {Frawley}}, \bibinfo {author} {\bibfnamefont {N.~R.}\ \bibnamefont
  {Fletcher}}, \ and\ \bibinfo {author} {\bibfnamefont {L.~C.}\ \bibnamefont
  {Dennis}},\ }\href {\doibase 10.1103/PhysRevC.25.860} {\bibfield  {journal}
  {\bibinfo  {journal} {Phys. Rev. C}\ }\textbf {\bibinfo {volume} {25}},\
  \bibinfo {pages} {860} (\bibinfo {year} {1982})}\BibitemShut {NoStop}%
\bibitem [{\citenamefont {Hudan}\ \emph {et~al.}(2023)\citenamefont {Hudan},
  \citenamefont {Johnstone}, \citenamefont {Kumar}, \citenamefont {deSouza},
  \citenamefont {Allen}, \citenamefont {Bardayan}, \citenamefont {Blankstein},
  \citenamefont {Boomershine}, \citenamefont {Carmichael}, \citenamefont
  {Clark}, \citenamefont {Coil}, \citenamefont {Henderson}, \citenamefont
  {O'Malley},\ and\ \citenamefont {von Seeger}}]{Hudan23}%
  \BibitemOpen
  \bibfield  {author} {\bibinfo {author} {\bibfnamefont {S.}~\bibnamefont
  {Hudan}}, \bibinfo {author} {\bibfnamefont {J.~E.}\ \bibnamefont
  {Johnstone}}, \bibinfo {author} {\bibfnamefont {R.}~\bibnamefont {Kumar}},
  \bibinfo {author} {\bibfnamefont {R.~T.}\ \bibnamefont {deSouza}}, \bibinfo
  {author} {\bibfnamefont {J.}~\bibnamefont {Allen}}, \bibinfo {author}
  {\bibfnamefont {D.~W.}\ \bibnamefont {Bardayan}}, \bibinfo {author}
  {\bibfnamefont {D.}~\bibnamefont {Blankstein}}, \bibinfo {author}
  {\bibfnamefont {C.}~\bibnamefont {Boomershine}}, \bibinfo {author}
  {\bibfnamefont {S.}~\bibnamefont {Carmichael}}, \bibinfo {author}
  {\bibfnamefont {A.}~\bibnamefont {Clark}}, \bibinfo {author} {\bibfnamefont
  {S.}~\bibnamefont {Coil}}, \bibinfo {author} {\bibfnamefont {S.~L.}\
  \bibnamefont {Henderson}}, \bibinfo {author} {\bibfnamefont {P.~D.}\
  \bibnamefont {O'Malley}}, \ and\ \bibinfo {author} {\bibfnamefont {W.~W.}\
  \bibnamefont {von Seeger}},\ }\href {\doibase 10.1103/PhysRevC.107.064612}
  {\bibfield  {journal} {\bibinfo  {journal} {Phys. Rev. C}\ }\textbf {\bibinfo
  {volume} {107}},\ \bibinfo {pages} {064612} (\bibinfo {year}
  {2023})}\BibitemShut {NoStop}%
\bibitem [{\citenamefont {Tighe}\ \emph {et~al.}(1993)\citenamefont {Tighe},
  \citenamefont {Kolata}, \citenamefont {Belbot},\ and\ \citenamefont
  {E.F.}}]{Tighe93}%
  \BibitemOpen
  \bibfield  {author} {\bibinfo {author} {\bibfnamefont {R.}~\bibnamefont
  {Tighe}}, \bibinfo {author} {\bibfnamefont {J.}~\bibnamefont {Kolata}},
  \bibinfo {author} {\bibfnamefont {M.}~\bibnamefont {Belbot}}, \ and\ \bibinfo
  {author} {\bibfnamefont {A.}~\bibnamefont {E.F.}},\ }\href {\doibase
  10.1103/PhysRevC.47.2699} {\bibfield  {journal} {\bibinfo  {journal} {Phys.
  Rev. C}\ }\textbf {\bibinfo {volume} {47}},\ \bibinfo {pages} {2699}
  (\bibinfo {year} {1993})}\BibitemShut {NoStop}%
\bibitem [{\citenamefont {A.~Hertz}\ \emph {et~al.}(1978)\citenamefont
  {A.~Hertz}, \citenamefont {Essel}, \citenamefont {Körner}, \citenamefont
  {Rehm},\ and\ \citenamefont {Sperr}}]{Hertz78}%
  \BibitemOpen
  \bibfield  {author} {\bibinfo {author} {\bibfnamefont {A.}~\bibnamefont
  {A.~Hertz}}, \bibinfo {author} {\bibfnamefont {H.}~\bibnamefont {Essel}},
  \bibinfo {author} {\bibfnamefont {H.~J.}\ \bibnamefont {Körner}}, \bibinfo
  {author} {\bibfnamefont {K.~E.}\ \bibnamefont {Rehm}}, \ and\ \bibinfo
  {author} {\bibfnamefont {P.}~\bibnamefont {Sperr}},\ }\href {\doibase
  10.1103/PhysRevC.18.2780} {\bibfield  {journal} {\bibinfo  {journal} {Phys.
  Rev. C}\ }\textbf {\bibinfo {volume} {18}},\ \bibinfo {pages} {2780(R)}
  (\bibinfo {year} {1978})}\BibitemShut {NoStop}%
\bibitem [{\citenamefont {Asher}\ \emph {et~al.}(2021)\citenamefont {Asher},
  \citenamefont {Almaraz-Calderon}, \citenamefont {Tripathi}, \citenamefont
  {Kemper}, \citenamefont {Baby}, \citenamefont {Gerken}, \citenamefont
  {Lopez-Saavedra}, \citenamefont {Morelock}, \citenamefont {Perello}, ,\ and\
  \citenamefont {Wiedenh{\"o}ver}}]{Asher21}%
  \BibitemOpen
  \bibfield  {author} {\bibinfo {author} {\bibfnamefont {B.~W.}\ \bibnamefont
  {Asher}}, \bibinfo {author} {\bibfnamefont {S.}~\bibnamefont
  {Almaraz-Calderon}}, \bibinfo {author} {\bibfnamefont {V.}~\bibnamefont
  {Tripathi}}, \bibinfo {author} {\bibfnamefont {K.~W.}\ \bibnamefont
  {Kemper}}, \bibinfo {author} {\bibfnamefont {L.~T.}\ \bibnamefont {Baby}},
  \bibinfo {author} {\bibfnamefont {N.}~\bibnamefont {Gerken}}, \bibinfo
  {author} {\bibfnamefont {E.}~\bibnamefont {Lopez-Saavedra}}, \bibinfo
  {author} {\bibfnamefont {A.~B.}\ \bibnamefont {Morelock}}, \bibinfo {author}
  {\bibfnamefont {J.~F.}\ \bibnamefont {Perello}}, , \ and\ \bibinfo {author}
  {\bibfnamefont {I.}~\bibnamefont {Wiedenh{\"o}ver}},\ }\href {\doibase
  10.1103/PhysRevC.103.044615} {\bibfield  {journal} {\bibinfo  {journal}
  {Phys. Rev. C}\ }\textbf {\bibinfo {volume} {103}},\ \bibinfo {pages}
  {044615} (\bibinfo {year} {2021})}\BibitemShut {NoStop}%
\bibitem [{\citenamefont {Gasques}\ \emph {et~al.}(2004)\citenamefont
  {Gasques}, \citenamefont {Chamon}, \citenamefont {Pereira}, \citenamefont
  {Alvarez}, \citenamefont {Rossi}, \citenamefont {Silva},\ and\ \citenamefont
  {Carlson}}]{Gasques04}%
  \BibitemOpen
  \bibfield  {author} {\bibinfo {author} {\bibfnamefont {L.~R.}\ \bibnamefont
  {Gasques}}, \bibinfo {author} {\bibfnamefont {L.~C.}\ \bibnamefont {Chamon}},
  \bibinfo {author} {\bibfnamefont {D.}~\bibnamefont {Pereira}}, \bibinfo
  {author} {\bibfnamefont {M.~A.~G.}\ \bibnamefont {Alvarez}}, \bibinfo
  {author} {\bibfnamefont {E.~S.}\ \bibnamefont {Rossi}}, \bibinfo {author}
  {\bibfnamefont {C.~P.}\ \bibnamefont {Silva}}, \ and\ \bibinfo {author}
  {\bibfnamefont {B.~V.}\ \bibnamefont {Carlson}},\ }\href {\doibase
  10.1103/PhysRevC.69.034603} {\bibfield  {journal} {\bibinfo  {journal} {Phys.
  Rev. C}\ }\textbf {\bibinfo {volume} {69}},\ \bibinfo {pages} {034603}
  (\bibinfo {year} {2004})}\BibitemShut {NoStop}%
\bibitem [{\citenamefont {Ring}(1996)}]{Ring96}%
  \BibitemOpen
  \bibfield  {author} {\bibinfo {author} {\bibfnamefont {P.}~\bibnamefont
  {Ring}},\ }\href {\doibase 10.1016/0146-6410(96)00054-3} {\bibfield
  {journal} {\bibinfo  {journal} {Prog. Part. Nucl. Phys.}\ }\textbf {\bibinfo
  {volume} {37}},\ \bibinfo {pages} {193} (\bibinfo {year} {1996})}\BibitemShut
  {NoStop}%
\bibitem [{\citenamefont {Serot}\ and\ \citenamefont
  {Walecka}(1986)}]{Serot86}%
  \BibitemOpen
  \bibfield  {author} {\bibinfo {author} {\bibfnamefont {B.~D.}\ \bibnamefont
  {Serot}}\ and\ \bibinfo {author} {\bibfnamefont {J.~D.}\ \bibnamefont
  {Walecka}},\ }\href@noop {} {\bibfield  {journal} {\bibinfo  {journal} {Adv.
  Nucl. Phys.}\ }\textbf {\bibinfo {volume} {16}},\ \bibinfo {pages} {1}
  (\bibinfo {year} {1986})}\BibitemShut {NoStop}%
\bibitem [{\citenamefont {Montagnoli}\ and\ \citenamefont
  {Stefanini}(2017)}]{Montagnoli17}%
  \BibitemOpen
  \bibfield  {author} {\bibinfo {author} {\bibfnamefont {G.}~\bibnamefont
  {Montagnoli}}\ and\ \bibinfo {author} {\bibfnamefont {A.}~\bibnamefont
  {Stefanini}},\ }\href@noop {} {\bibfield  {journal} {\bibinfo  {journal}
  {Eur. Phys. J. A}\ }\textbf {\bibinfo {volume} {53}},\ \bibinfo {pages} {169}
  (\bibinfo {year} {2017})}\BibitemShut {NoStop}%
\bibitem [{\citenamefont {Dasgupta}\ \emph {et~al.}(1998)\citenamefont
  {Dasgupta}, \citenamefont {Hinde}, \citenamefont {Rowley},\ and\
  \citenamefont {Stefanini}}]{DasGupta98}%
  \BibitemOpen
  \bibfield  {author} {\bibinfo {author} {\bibfnamefont {M.}~\bibnamefont
  {Dasgupta}}, \bibinfo {author} {\bibfnamefont {D.~J.}\ \bibnamefont {Hinde}},
  \bibinfo {author} {\bibfnamefont {N.}~\bibnamefont {Rowley}}, \ and\ \bibinfo
  {author} {\bibfnamefont {A.~M.}\ \bibnamefont {Stefanini}},\ }\href {\doibase
  doi.org/10.1146/annurev.nucl.48.1.401} {\bibfield  {journal} {\bibinfo
  {journal} {Ann. Rev. Nucl. Part. Sci.}\ }\textbf {\bibinfo {volume} {48}},\
  \bibinfo {pages} {401} (\bibinfo {year} {1998})}\BibitemShut {NoStop}%
\end{thebibliography}
%

\end{document}